\documentclass[twocolumn,floatfix,showpacs,prb]{revtex4}
\usepackage{graphicx}
\usepackage{amsmath}
\usepackage{bm}
\usepackage{psfrag}
\begin{document}

\title{Ballistic transmission through a graphene bilayer}
\author{I. Snyman}
\affiliation{Instituut-Lorentz, Universiteit Leiden, P.O. Box 9506, 2300 RA Leiden, The Netherlands}
\author{C. W. J. Beenakker}
\affiliation{Instituut-Lorentz, Universiteit Leiden, P.O. Box 9506, 2300 RA Leiden, The Netherlands}
\date{September 2006}
\begin{abstract}
We calculate the Fermi energy dependence of the (time-averaged) current and shot noise in an impurity-free carbon bilayer (length $L\ll$ width $W$), and compare with known results for a monolayer. At the Dirac point of charge neutrality, the bilayer transmits as two independent monolayers in parallel: Both current and noise are resonant at twice the monolayer value, so that their ratio (the Fano factor) has the same $1/3$ value as in a monolayer --- and the same value as in a diffusive metal. The range of Fermi energies around the Dirac point within which this pseudo-diffusive result holds is smaller, however, in a bilayer than in a monolayer (by a factor $l_{\perp}/L$, with $l_{\perp}$ the interlayer coupling length).
\end{abstract}
\maketitle

\section{Introduction}
\label{intro}
Undoped graphene has no free electrons, so an infinite sample cannot conduct electricity. A finite sample can conduct, because electrons
injected at one end can be transmitted a distance $L$ to the other end via so-called evanescent modes. These are modes that decay 
$\propto e^{-L/\lambda}$ with a penetration depth $\lambda$ bounded from above by the width $W$ of the sample. For a wide and narrow sample
($W\gg L$), there are many evanescent modes that contribute appreciably to the conductance.
Because the transmission of an electron via an evanescent mode is a stochastic event, the current fluctuates in time --- even in the absence of any scattering by impurities or lattice defects. Tworzyd{\l}o  et al.\cite{Two06} found that the shot noise produced by the evanescent modes in an undoped carbon monolayer (of length $L$ $\ll$ width $W$) is {\em pseudo-diffusive}: The Fano factor $F=P/2e\bar{I}$ (ratio of noise power $P$ and time-averaged current $\bar{I}$) has the same value $F=1/3$ as in a diffusive metal (while $F=1$ for independent current pulses).\cite{reviews}

A carbon bilayer has an additional length scale, not present in the monolayer of Ref.\ \onlinecite{Two06}, namely the interlayer coupling length $l_{\perp}$. It is an order of magnitude larger than the interatomic distance $d$ within the layer:\cite{Wal47,McCann06,Nil06b}
\begin{equation}
l_{\perp}=\frac{\hbar v}{t_{\perp}}=\frac{3t_{\parallel}}{2t_{\perp}}\,d\approx 11\,d \label{lperpdef}
\end{equation}
(with $v\approx 10^6\,{\rm m/s}$, $d\approx 1.4\,\mbox{\AA}$, and $t_{\parallel}\approx 3\,{\rm eV}$, respectively the carrier velocity, interatomic distance, and nearest-neigbour hopping energy within a single layer, and $t_{\perp}\approx 0.4\,{\rm eV}$ the nearest-neighbour hopping energy between two layers\cite{note1}). Since $L$ is typically large compared to $l_{\perp}$, the two layers are strongly coupled. In this paper we investigate what is the effect of interlayer coupling on the average current and shot noise.

The model and calculation are outlined in Secs.\ \ref{model} and \ref{transmission}. Our main conclusion, presented in Sec.\ \ref{results}, is that an undoped graphene bilayer has the same current and noise as two monolayers in parallel. The Fano factor, therefore, still equals $1/3$ when the Fermi level coincides with the Dirac point (at which conduction and valence bands touch). However, the interval $\Delta E_F\simeq\hbar v l_{\perp}/L^{2}$ in Fermi energy around the Dirac point where this pseudo-diffusive result holds is much narrower, by a factor $l_{\perp}/L$, in a bilayer than it is in a monolayer.

Our results for the mean current $\bar{I}$, and hence for the conductance in a ballistic system, agree with those of Cserti,\cite{Cse06} but differ from two other recent calculations in a (weakly) disordered system.\cite{Kos06,Nil06} (The shot noise was not considered in Refs.\ \onlinecite{Cse06,Kos06,Nil06}.) A ballistic system like ours was studied recently by Katsnelson,\cite{Kat06b} with different results for both conductance and shot noise. We discuss the origin of the difference in Sec.\ \ref{potentialdependence}. We conclude by connecting with experiments\cite{Nov06} in Sec.\ \ref{conclusion}.

\section{Model}
\label{model}

\begin{figure}[tb]
\vspace*{0.1\linewidth}

\centerline{\includegraphics[width=0.9\linewidth]{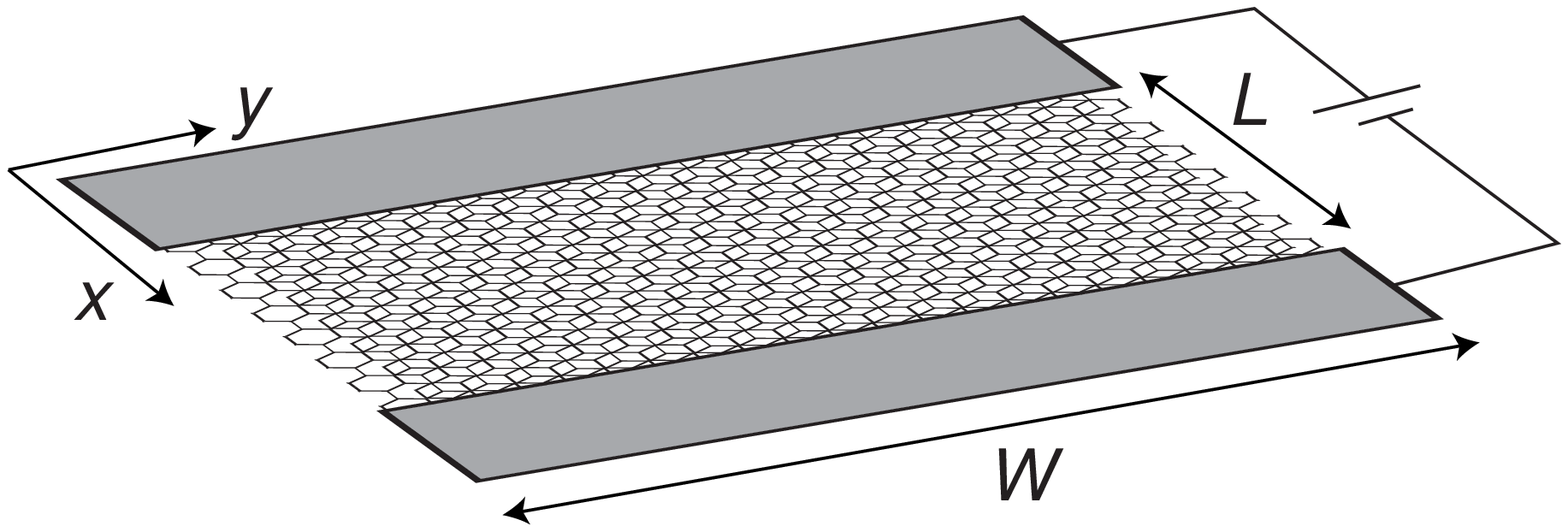}}\medskip

\centerline{\includegraphics[width=0.9\linewidth]{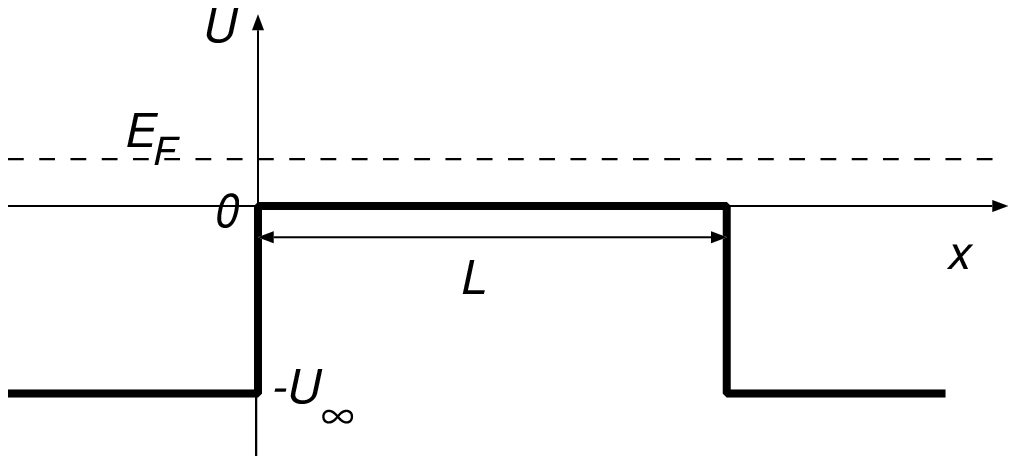}}
\caption{\label{fig_bilayer}
Schematic of the graphene bilayer. Top panel: Two stacked honeycomb lattices of carbon atoms in a strip between metal contacts. Bottom panel: Variation of the electrostatic potential across the strip.}
\end{figure}

We use the same setup as in Refs.\ \onlinecite{Two06,Kat06}, shown schematically in Fig.\ \ref{fig_bilayer}. A sheet of ballistic graphene in the $x-y$ plane contains a weakly doped strip of width $W$ and length $L$, and heavily doped contact regions for $x<0$ and $x>L$. The doping is controlled by gate voltages, which induce a potential profile of the form
\begin{equation}
U(x)=\left\{\begin{array}{cl}
-U_{\infty}&{\rm if}\;\;x<0\;\;{\rm or}\;\;x>L,\\
0&{\rm if}\;\;0<x<L.
\end{array}\right.\label{Udef}
\end{equation}
We use an abrupt potential step for simplicity, justified by the fact that any smoothing of the step over a distance small compared to $L$ becomes irrelevant near the Dirac point, when the Fermi wave length $\gtrsim L$.

While Refs.\ \onlinecite{Two06,Kat06} considered a graphene monolayer, governed by the $2\times 2$ Dirac Hamiltonian, here we take a bilayer with $4\times 4$ Hamiltonian\cite{Wal47,McCann06,Nil06b}
\begin{equation}
H=\begin{pmatrix}
U&v(p_{x}+ip_{y})&t_{\perp}&0\\
v(p_{x}-ip_{y})&U&0&0\\
t_{\perp}&0&U&v(p_{x}-ip_{y})\\
0&0&v(p_{x}+ip_{y})&U
\end{pmatrix},\label{Hdef}
\end{equation}
with $\bm{p}=-i\hbar\partial/\partial{\bm r}$ the momentum operator. The Hamiltonian acts on a four-component spinor $(\Psi_{A_{1}},\Psi_{B_{1}},\Psi_{B_{2}},\Psi_{A_{2}})$ with amplitudes on the $A$ and $B$ sublattices of the first and second layer. Only nearest-neighbor hopping is taken into account, either from $A$ to $B$ sites within a layer or between different layers. (Sites from the same sublattice but on different layers are not directly adjacent.) The Hamiltonian (\ref{Hdef}) describes low-energy excitations near one of the two Dirac points in the Brillouin zone, where conduction and valence bands touch. The other Dirac point and the spin degree of freedom contribute a four-fold degeneracy factor to current and noise power.

We have taken the same electrostatic potential $U$ in both layers. In general, the potentials will differ,\cite{Nil06c,McC06} but to study the special physics of undoped graphene it is necessary that they are both tuned to the Dirac point of each layer. This can be achieved by separate top and bottom gates (not shown in Fig.\ \ref{fig_bilayer}).

\begin{figure}[t]
  \begin{center}
    \includegraphics[width=.9 \columnwidth]{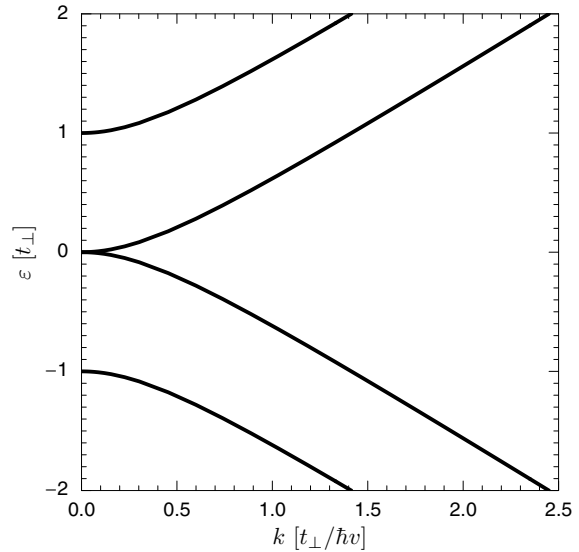}
    \caption{Energy spectrum (\ref{bilayerspectrum}) of the graphene bilayer, according to the Hamiltonian (\ref{Hdef}).
  \label{fig1}}
  \end{center}
\end{figure}

For free electrons in bilayer graphene, the relation between energy $\varepsilon$ and total momentum $k=(k_x^2+k_y^2)^{1/2}$ as described by this 
Hamiltonian consists of four hyperbolas, defined by
\begin{subequations}
\label{bilayerspectrum}
\begin{align}
\varepsilon&=\pm\tfrac{1}{2}t_\perp\pm \sqrt{\tfrac{1}{4}t_\perp^2+k^2},\label{spectruma}\\
\varepsilon&=\mp\tfrac{1}{2}t_\perp\pm \sqrt{\tfrac{1}{4}t_\perp^2+k^2},\label{spectrumb}
\end{align}
\end{subequations}
plotted in Fig.\ \ref{fig1}. (For notational convenience, we use units such that $\hbar v=1$ in most equations.) 

We calculate the transmission matrix $\bm t$ through the graphene strip at the Fermi energy, 
and then obtain the conductance and noise power from the Landauer-B\"{u}ttiker formulas\cite{reviews}
\begin{eqnarray}
&&G=G_{0}\,{\rm Tr}\,\bm{tt}^{\dagger},\;\;P=P_{0}\,{\rm Tr}\,\bm{tt}^{\dagger}(1-\bm{tt}^{\dagger}),\label{GPdef}\\
&&\rightarrow F=\frac{{\rm Tr}\,\bm{tt}^{\dagger}(1-\bm{tt}^{\dagger})}{{\rm Tr}\,\bm{tt}^{\dagger}},\label{Fdef}
\end{eqnarray}
with $G_{0}=4e^{2}/h$, $P_{0}=2e|V|G_{0}$, and $V$ the voltage applied between the contact regions. The results depend on the degree of doping in the graphene strip (varied by varying $E_{F}$), but they become independent of the degree of doping of the contact regions if $U_{\infty}\gg t_{\perp}$.

\section{Transmission probabilities}
\label{transmission}

We calculate the transmission matrix by matching eigenstates of the Hamiltonian (\ref{Hdef}) at the two interfaces $x=0$ and $x=L$. 
This procedure is similar to a calculation of non-relativistic scattering by a rectangular barrier in a two-dimensional waveguide. 
There are two differences. Firstly, the Hamiltonian (\ref{Hdef}) is a first-order differential operator, and hence 
only the wavefunction and not its derivative is continuous at the interface. Secondly, the
spectrum contains both positive and negative energy eigenstates.

The eigenstates of $H$ for $U=0$ have been given in Ref.\ \onlinecite{Nil06c}.  
They may be characterized as follows. For given energy $\varepsilon$ and
transverse momentum $k_y$, we define two longitudinal momenta
\begin{equation}
k_{x\pm}=\sqrt{(\varepsilon\pm \tfrac{1}{2}t_\perp)^2-\tfrac{1}{4}t_\perp^2-k_y^2}.
\end{equation}
The square root is taken with argument in the interval $[0,\pi)$. Associated with each real $k_{x+}$ there are two propagating modes, one left-going $\phi_{\varepsilon,+}^{L}$ and one right-going $\phi_{\varepsilon,+}^{R}$. Two more propagating modes $\phi_{\varepsilon,-}^{L}$ and $\phi_{\varepsilon,-}^{R}$ are associated with each
real $k_{x-}$. These eigenstates of $H$ are given by
\begin{subequations}
\begin{align}
\phi_{\varepsilon,\pm}^R(x,y)&=N_\pm\left(\begin{array}{c}\mp \varepsilon\\\mp k_{x\pm}\pm ik_y\\\varepsilon
\\k_{x\pm}+ik_y\end{array}\right)e^{i(k_{x\pm}x+k_yy)},\\
\phi_{\varepsilon,\pm}^L(x,y)&=N_\pm\left(\begin{array}{c}\mp \varepsilon\\\pm k_{x\pm}\pm ik_y\\
\varepsilon\\-k_{x\pm}+ik_y\end{array}\right)e^{i(-k_{x\pm}x+k_yy)},
\end{align}
\end{subequations}
with $N_{\pm}=(4 W \varepsilon k_{x\pm})^{-\tfrac{1}{2}}$ a normalization constant such that 
each state carries unit current
\begin{equation}
I=ev\int_0^W dy\ \phi^\dagger\left(\begin{array}{cc}\sigma_x&0\\0&\sigma_x\end{array}\right)\phi,
\end{equation}
in the positive or negative $x$-direction.

For each $k_{y}$ we have two left-incident scattering states $\psi_{\varepsilon,\pm}$ at energy $\varepsilon$. In the region $x<0$ to the left of the strip they have the form
\begin{equation}
\psi_{\varepsilon,\pm}=\phi_{\varepsilon+U_\infty,\pm}^{R}+
r^\pm_+(\varepsilon,k_y)\phi_{\varepsilon+U_\infty,+}^L+r^\pm_-(\varepsilon,k_y)\phi_{\varepsilon+U_\infty,-}^L,\label{eq3a}
\end{equation}
while to the right of the strip ($x>L$) one has
\begin{equation}
\psi_{\varepsilon,\pm}=
t^\pm_+(\varepsilon,k_y)\phi_{\varepsilon+U_\infty,+}^R+t^\pm_-(\varepsilon,k_y)\phi_{\varepsilon+U_\infty,-}^R.\label{eq3b}
\end{equation}
For $\varepsilon\not=0$ the form of the solution in the region $x\in[0,L]$ is self-evidently a linear combination of the four solutions $\phi_{\varepsilon\pm}^{L}$, $\phi_{\varepsilon\pm}^{R}$. Care must however be taken in analytical work to use proper linear combinations of these modes that remain linearly independent exactly at $\varepsilon=0$ (the Dirac point). (See Appendix \ref{detail} for explicit formulas.)

The four transmission amplitudes $t_\pm^\pm$ for given $\varepsilon$ and $k_{y}$ can be combined in the transmission matrix
\begin{equation}
\bm t(\varepsilon,k_y)=\left(\begin{array}{cc}t_+^+(\varepsilon,k_y)&t_-^+(\varepsilon,k_y)
\\t_+^-(\varepsilon,k_y)&t_-^-(\varepsilon,k_y)\end{array}\right).
\end{equation}
We consider a short and wide geometry $L\ll W$, in which the boundary conditions in the $y$-direction become irrelevant. For simplicity, we take periodic boundary conditions, such that $k_{y}$ is quantized as $k_{y,n}=2\pi n/W$, $n=0,\pm 1,\pm 2,\ldots$. In the regime $L\ll W$, $|\varepsilon|\ll U_{\infty}$ considered here, both the discreteness and the finiteness of the modes in the contact region can be ignored. As a consequence, the traces in Eqs.\ (\ref{GPdef}) and (\ref{Fdef}) may be replaced by integrals through the prescription
\begin{equation}
{\rm Tr}\,(\bm{tt}^{\dagger})^{p}\rightarrow\frac{W}{\pi}\int_0^\infty dk_y\,\sum_{\sigma=\pm}\left[T_{\sigma}(E_F,k_{y})\right]^{p},
\label{sumint}
\end{equation}
where $T_\pm$ are the two eigenvalues of $\bm{tt^\dagger}$.

\section{Results}
\label{results}
\begin{figure}[t]
  \begin{center}
    \includegraphics[width=.9 \columnwidth]{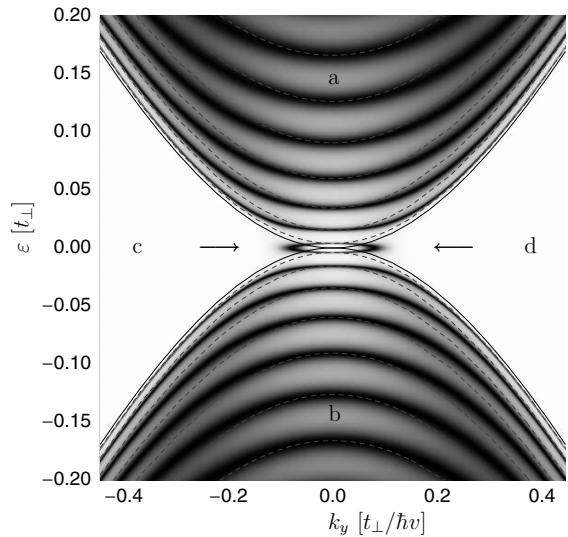}
    \caption{Total transmission probability ${\rm Tr}(\bm{tt^\dagger})$ as a function of $\varepsilon$ and $k_y$ 
    for $U_\infty=50\,t_\perp$ and $L=50\,l_\perp$. Darkly
    shaded regions indicate high transmission. Grey dashed lines indicate the estimate (\ref{eqa}) for the occurrence of 
    resonances in regions (a) and (b), while solid lines indicate the boundary between propagating and
    evanescent modes. Arrows point to the resonances of evanescent modes close to the Dirac point, responsible for the 
    pseudo-diffusive transport.
  \label{fig2}}
  \end{center}
\end{figure}
Fig.\ \ref{fig2} contains a grey-scale plot of the total transmission probability ${\rm Tr}(\bm{tt^\dagger})$ 
as a function of $k_y$ and $\varepsilon$.
Darkly shaded regions indicate resonances of high transmission, similar to those found in Ref.\ \onlinecite{Kat06c}. 

The location $\varepsilon_{\rm res}$
of resonances can be estimated by equating $k_x L/\pi$ to an integer $n$. 
This yields the curves
\begin{equation}
\varepsilon^{(n)}_{\rm res}(k_y)=\mp \tfrac{1}{2}t_\perp\pm\sqrt{\tfrac{1}{4}t_\perp^2+\left(\frac{\pi n}{L}\right)^2+k_y^2},\label{eqa}
\end{equation}
indicated in the figure by dashed lines. It is seen that good agreement is reached for $|k_y|\ll 1/L$ and again for
$|k_y|\gg1/L$. For $|k_y L|\simeq 1$ there is a cross over. 
In regions (c) and (d), demarkated by the curves $\varepsilon^{(0)}_{\rm res}$, the
transmission generally drops to zero, since in these regions the longitudinal momentum $k_x$ is imaginary. 

There is however a
curious feature close to $\varepsilon,k_y=0$. The resonance closest to the Dirac point 
behaves differently from all the other resonances. When $|k_y|$ is increased, it moves closer to the Dirac point rather than away from it,
eventually crossing into regions (c) and (d) of evanescent modes. 
It is this resonance of evanescent modes that is responsible for the pseudo-diffusive transport at the Dirac point. 

\begin{figure}[t]
  \begin{center}
    \includegraphics[width=.9 \columnwidth]{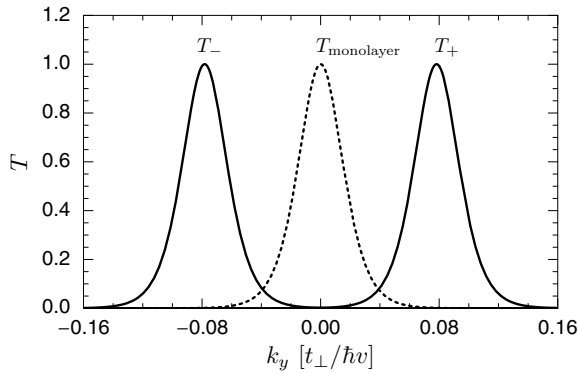}
    \caption{Solid curves: Transmission coefficients $T_\pm$ of the bilayer according to Eq.\ (\ref{eq7}) at $L=50\,l_\perp$. These
    coefficients are displaced copies of the monolayer result (dashed).\label{t0}}
  \end{center}
\end{figure}

At $\varepsilon=0$, 
the exact formula for
the eigenvalues of $\bm{tt^\dagger}$ in the $U_\infty\rightarrow\infty$ limit is
\begin{align}
&T_\pm(\varepsilon=0,k_y)=\frac{1}{\cosh^2(k_y\mp k_c) L},\label{eq7}\\
&k_c=\frac{1}{L}\ln\left[\frac{L}{2l_\perp}+\sqrt{1+\frac{L^2}{4l_\perp^2}}\ \right].
\end{align}
In Fig.\ \ref{t0} the two transmission coefficients $T_\pm(0,k_y)$ are compared to the single transmission coefficient
$T_{\rm monolayer} (0,k_y)=1/\cosh^2(k_yL)$ of the monolayer.\cite{Two06,Kat06}
Details of the calculation may be found in Appendix \ref{detail}.

Since the two bilayer coefficients are displaced copies of the monolayer coefficient, any observable of the form 
$A={\rm Tr}\ f(\bm{tt^\dagger})$, with $f$ an arbitrary function is twice as large in a bilayer as it is in a monolayer.
From Eqs. (\ref{GPdef}) and (\ref{sumint}) we obtain
\begin{eqnarray}
&&G_{\rm bilayer}=2G_{\rm monolayer}=\frac{2G_0}{\pi}\frac{W}{L},\label{Gbi}\\
&&P_{\rm bilayer}=2P_{\rm monolayer}=\frac{4e|V|G_0}{3\pi}\frac{W}{L} \label{Pbi},\\
&&F_{\rm bilayer}=F_{\rm monolayer}=\tfrac{1}{3}.
\end{eqnarray}

\begin{figure}[t]
  \begin{center}
    \includegraphics[width=.9 \columnwidth]{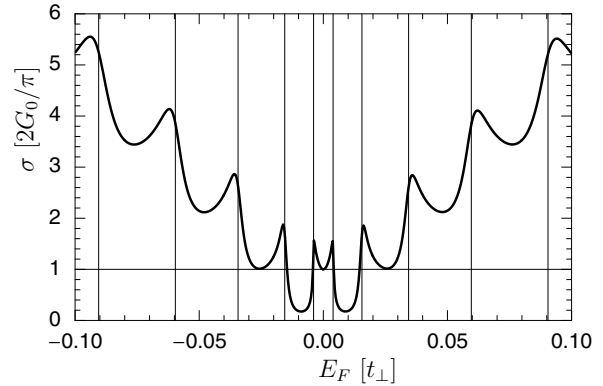}
  \end{center}
  \begin{center}
    \includegraphics[width=.9 \columnwidth]{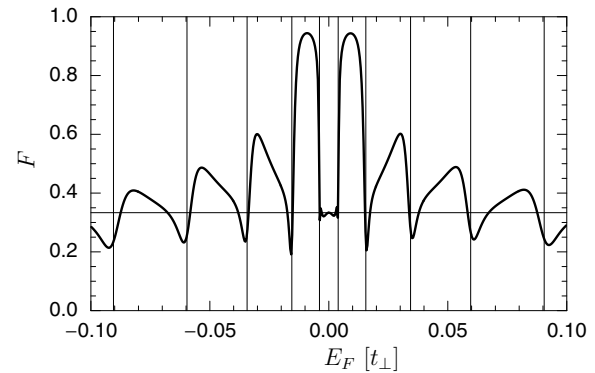}
    \caption{Conductivity $\sigma$ (top) and Fano factor $F$ (bottom) of the bilayer,
    as a function of the Fermi energy $E_F$ measured from the Dirac point 
    for $U_\infty=50\,t_\perp$ and $L=50\,l_\perp$. Abrupt features occur at 
    $E_F\simeq\varepsilon_{\rm res}^{(n)}(k_y=0)$ [vertical lines, given by Eq. (\ref{res})].\label{fig4}}
  \end{center}
\end{figure}
Figure \ref{fig4} contains plots of both the conductivity $\sigma=GL/W$ and Fano factor of the bilayer around the 
Dirac point. At energies associated with resonances at normal incidence,
\begin{equation}
\varepsilon_{\rm res}^{(n)}(0)=\pm \frac{\pi^2\hbar v}{L}\left[\frac{l_\perp}{L}n^2+\mathcal O(l_\perp/L)^3\right],\label{res}
\end{equation}
the conductivity and Fano factor show abrupt features. The width $\Delta E_F=2\varepsilon_{\rm res}^{(1)}=2\pi^2 \hbar vl_\perp/L^2$
of the energy window between the resonances that straddle the Dirac point in the bilayer is smaller by a factor $l_\perp/L$ than
in the monolayer.

\section{Dependence on the potential in the contact region}
\label{potentialdependence}

So far we have assumed that the potential $U_\infty$ in the contact region is large compared to the band splitting $t_\perp$ near the Dirac point 
of the graphene bilayer. 
We believe that this is the appropriate regime to model a normal metal contact to the graphene sheet, which couples equally 
well to the two sublattices on each layer.

\begin{figure}[tb]
  \begin{center}
    \vspace*{1mm}
    \includegraphics[width=.9 \columnwidth]{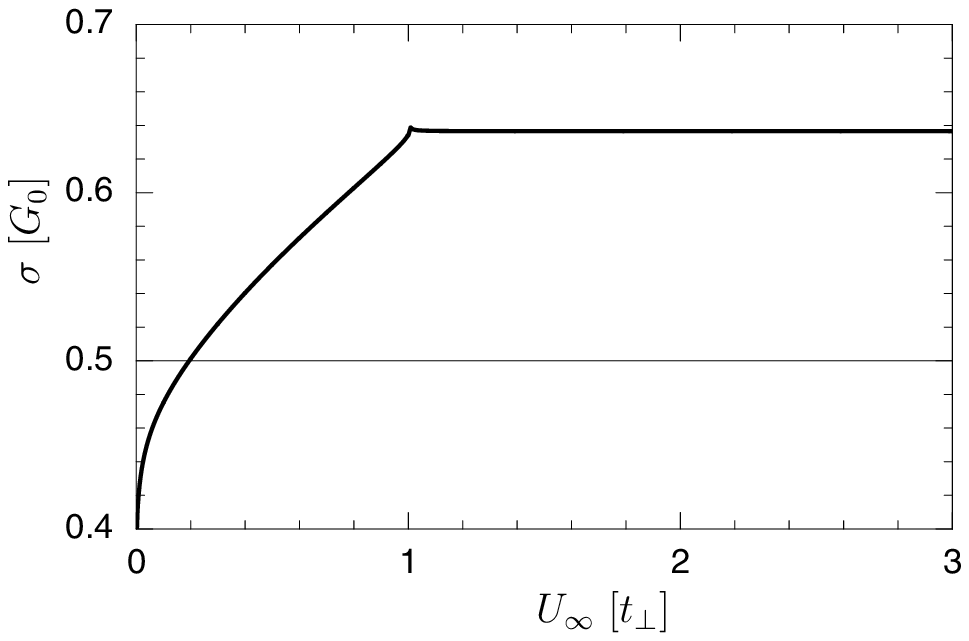}
  \end{center}
  \begin{center}
    \includegraphics[width=.9 \columnwidth]{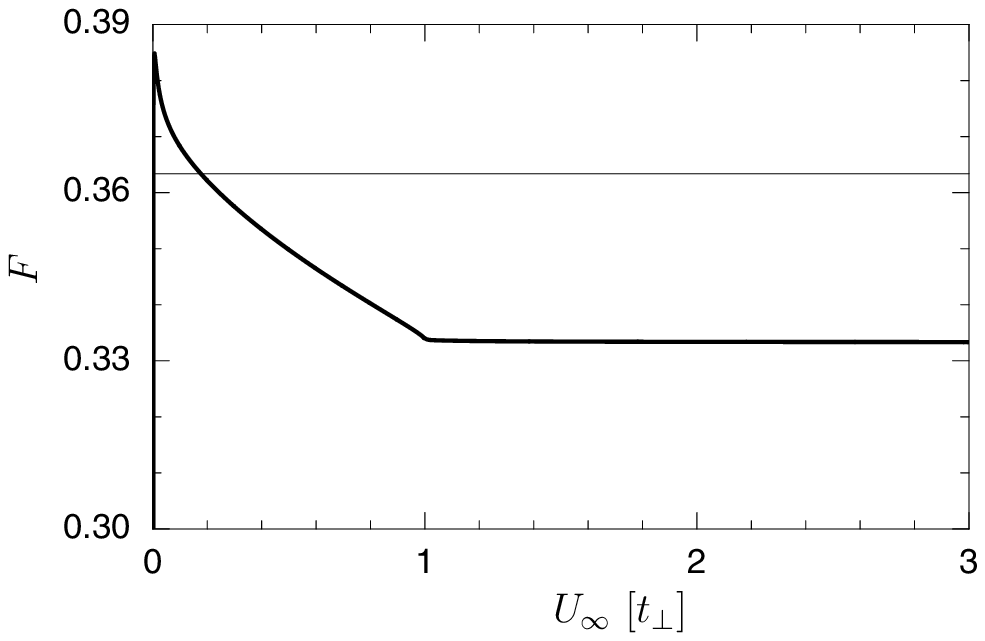}
    \caption{Dependence of the conductivity and Fano factor at the Dirac point on the potential $U_\infty$ in the contact region, for $L=100\ l_\perp$.
    Thin horizontal lines indicate the values of Ref.\ \onlinecite{Kat06b}. The values obtained in this paper correspond to a plateau reached
    for $U_\infty/t_\perp\gtrsim 1$. \label{cfv}}
  \end{center}
\end{figure}

It is of interest to determine how large the ratio $U_{\infty}/t_{\perp}$ should be to reach the contact-independent limit of the previous section.
Note that for $U_\infty>t_\perp$ there are two left-incident propagating modes in the leads for each $\varepsilon$ and $k_y$. When $U_\infty$ becomes smaller than $t_\perp$ one of the two modes becomes evanescent, leading to an abrupt change in the conductivity and the Fano factor. This is evident in Fig.\ \ref{cfv}. For $U_{\infty}-t_\perp\gtrsim \hbar v/L$, the conductivity and Fano factor have almost reached their $U_\infty\rightarrow\infty$ limits. For $U_\infty\lesssim t_\perp$ the conductivity is smaller and the Fano factor larger than when $U_\infty>t_\perp$. 
Both quantities vanish when the Fermi momentum 
$\sqrt{U_\infty t_\perp}/v$ in the contact region  drops below $\hbar /L$ and the contact region is effectively depleted of carriers.

These finite-$U_{\infty}$ results can be used to make contact with the previous calculation of Katsnelson,\cite{Kat06b} who found a conductivity
$\sigma=G_0/2$ and a Fano factor $F=1-2/\pi$ at the Dirac point, in the regime $\hbar v/L\ll \sqrt{U_\infty t_\perp}\ll t_\perp$. These values are indicated in Fig.\ \ref{cfv} by horizontal lines. The intersection point with our curves occurs at nearly the same value of $U_{\infty}/t_{\perp}$ for both quantities. 
The intersection point moves closer and closer to $U_\infty=0$ as the sample length $L$ is increased, 
but there is no clear plateau around the intersection point. 
Moreover, as shown in Appendix B, the intersection point does not correspond to a minimum or maximum as a function of the Fermi energy, so that these values would be difficult to extract from a measurement. 

We do believe that the results of Ref.\ \onlinecite{Kat06b} describe the asymptotic limit $L/l_\perp\rightarrow\infty$ at $E_{F}\equiv 0$. however, because in this limit the width $\Delta E_F\simeq \hbar v l_{\perp}/L^{2}$ of the resonance at the Dirac point vanishes, it seems unobservable.

\section{Conclusion}
\label{conclusion}

In conclusion, we have demonstrated that the pseudo-diffusive transport 
at the Dirac point, discovered in Ref.\ \onlinecite{Two06} for a carbon 
monolayer, holds in a bilayer as well. All moments of the current 
fluctuations have the same relation to the mean current as in a 
diffusive metal. In particular, the Fano factor has the $1/3$ value 
characteristic of diffusive transport, even though the bilayer is 
assumed to be free of impurities or lattice defects.

Although we found that an undoped bilayer transmits as two undoped 
monolayers in parallel, the two systems behave very different away from 
charge neutrality. The resonance of evanescent modes around the Dirac 
point of zero Fermi energy has width $\Delta E_{F}\simeq \hbar v 
l_{\perp}/L^{2}$ in a bilayer, which is smaller than the width in a 
monolayer by the ratio of the interlayer coupling length $l_{\perp}$ and 
the separation $L$ of the metal contacts.

Since $l_{\perp}\approx 1.5\,{\rm nm}$, one would not be able to resolve 
this resonance in the $\mu{\rm m}$-size samples of Ref.\ \onlinecite{Nov06}. 
These experiments found no qualitative difference in the 
conductance-versus-gate-voltage dependence of monolayer and bilayer 
graphene, both showing a minimum conductivity at the Dirac point of 
$G_{0}$. Smaller junctions in the 10--100~nm range as are now being 
fabricated should make it possible to resolve the transmission resonance 
of evanescent modes predicted here, and to observe the unusual 
pseudo-diffusive dynamics associated with it.

\acknowledgments

This research was supported by the Dutch Science Foundation NWO/FOM. We have benefitted from discussions with E. McCann.

\appendix
\begin{widetext}
\section{Transmission eigenvalues at the Dirac point}
\label{detail}
In this Appendix we give some detail of the calculation that leads to the transmission coefficients $T_\pm(\varepsilon=0,k_y)$ of Eq.\ (\ref{eq7}).
At the Dirac point and in the limit of large $U_\infty$, the left-incident eigenstates of the Hamiltonian (\ref{Hdef}) are of the form
\begin{equation}
\psi_\pm(x)=\left\{\begin{array}{ll}
\raisebox{2mm}{$e^{ik_y y}\left[\xi^R_{\pm}e^{iU_\infty x}+\left(r^\pm_+\xi^L_+ + r^\pm_-\xi^L_-\right)e^{-iU_\infty x}\right]$}&
\raisebox{2mm}{$\hspace{5mm}x<0$,}\\
e^{ik_y y}\left[\left(c^\pm_1\chi_1+c^\pm_2\chi_2\right)e^{k_y x}+\left(c^{\pm}_3\chi_3+c^{\pm}_4\chi_4\right)e^{-k_y x}\right]&\hspace{5mm}0<x<L,\\
\raisebox{-2mm}{$e^{ik_y y}e^{iU_\infty(x-L)}\left[t^\pm_+\xi^R_ + + t^\pm_-\xi^R_-\right]$}&\raisebox{-2mm}{$\hspace{5mm}x>L$,}\end{array}\right.
\end{equation}
with the definitions
\begin{equation}
\xi^R_\pm=\left(\begin{array}{c}\mp1\\\mp1\\1\\1\end{array}\right),\hspace{2mm} \xi^L_\pm=\left(\begin{array}{c}\mp1\\\pm1\\1\\-1\end{array}\right),
\end{equation}
\begin{equation}
\chi_{1}=\left(\begin{array}{c}0\\1\\0\\0\end{array}\right),\hspace{2mm} 
\chi_{2}=\left(\begin{array}{c}0\\-i t_\perp x\\1\\0\end{array}\right),
\chi_{3}=\left(\begin{array}{c}1\\0\\0\\-it_\perp x\end{array}\right),\hspace{2mm}
\chi_{4}=\left(\begin{array}{c}0\\0\\0\\1\end{array}\right),\hspace{2mm} 
\end{equation}
These eigenstates must be continuous at $x=0$ and $x=L$, leading to an $8\times8$ system of linear equations $M\bm{b_\pm}=\bm{c_\pm}$ with
\begin{equation}
M=\left(\begin{array}{rrrcccrr}
1 &- 1 & 0 & 0 & 1 & 0 & 0 & 0\\
-1 & 1 & 1 & 0 & 0 & 0 & 0 & 0\\
-1 & -1 & 0 & 0 & 0 & 1 & 0 & 0\\
1 & 1 & 0 & 1 & 0 & 0 & 0 & 0\\
0 & 0 & 0 & 0 & 1 & 0 & z & -z\\
0 & 0 & z & -iL t_\perp z & 0 & 0& 1 & -1\\
0 & 0 & 0 & z & 0 & 0 & -1 & -1\\
0 & 0 & 0 & 0 & -iLt_\perp & 1 & -z & -z\end{array}\right),\hspace{2mm}
\bm{b}_\pm=\left(\begin{array}{c}r_+^\pm\\r_-^\pm\\c_1\\c_2\\c_3\\c_4\\t_+^\pm\\t_-^\pm\end{array}\right),\hspace{2mm}
\bm{c}_\pm=\left(\begin{array}{l}\mp1\\\mp1\\1\\1\\0\\0\\0\\0\end{array}\right). 
\end{equation}
We abbreviated $z=e^{k_y L}$.
By solving these equations, one finds the transmission matrix
\begin{equation}
\bm{t}=\frac{2i}{2+\left(L/l_\perp\right)^2+2\cosh(2k_yL)}\left(\begin{array}{cc}
\left(L/l_\perp-2i\right)\cosh(k_yL)&(L/l_\perp)\sinh(k_yL)\\
-(L/l_\perp)\sinh(k_yL)&-\left(L/l_\perp+2i\right)\cosh(k_yL)\end{array}\right).
\end{equation}
The eigenvalues of $\bm{tt^\dagger}$ are then given by Eq.\ (\ref{eq7}).
\end{widetext}
\section{Four-band versus two-band Hamiltonian}
\label{fourvstwo}

In this Appendix we verify that the difference in the results obtained here and in Ref.\ \onlinecite{Kat06b} is not due to the different Hamiltonians used in these two calculations.

In Ref.\ \onlinecite{Kat06b} the limit $t_{\perp}\rightarrow\infty$ was taken at the beginning of the calculation, reducing the $4\times 4$ Hamiltonian (\ref{Hdef}) to the effective $2\times 2$ Hamiltonian\cite{McCann06}
\begin{equation}
H_{\rm eff}=-\frac{v^2}{t_\perp}\left(\begin{array}{cc}0&(p_x-ip_y)^2\\(p_x+ip_y)^2&0\end{array}\right)
+U(x)\left(\begin{array}{cc}1&0\\0&1\end{array}\right).\label{H2def}
\end{equation}
Only the two lowest bands near the Dirac point are retained in $H_{\rm eff}$, as is appropriate for the regime $U_{\infty}\ll t_{\perp}$.

\begin{figure}[tb]
  \begin{center}
    \includegraphics[width=.9 \columnwidth]{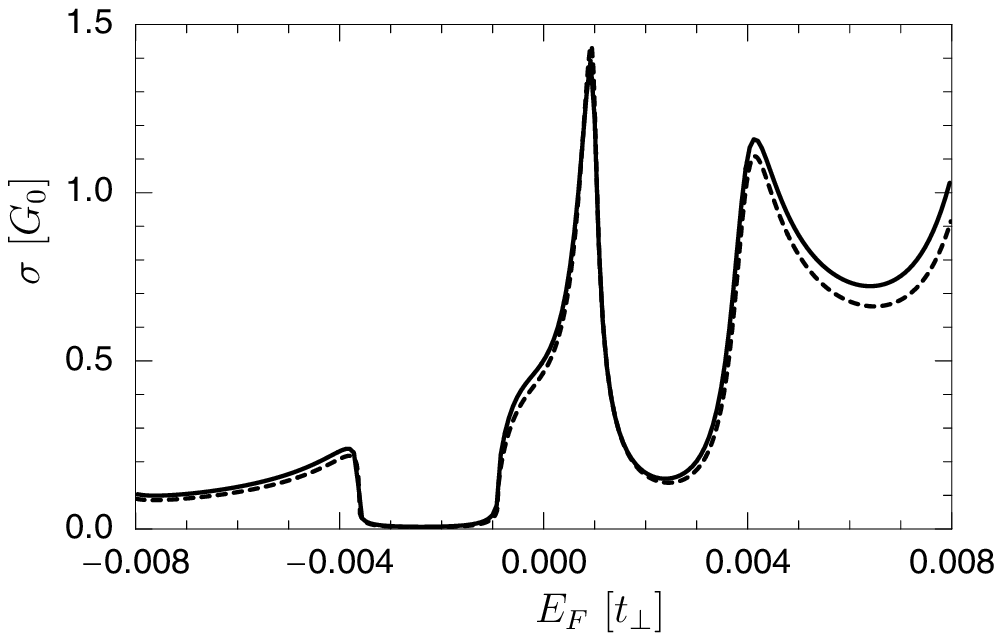}
  \end{center}
  \begin{center}
    \includegraphics[width=.9 \columnwidth]{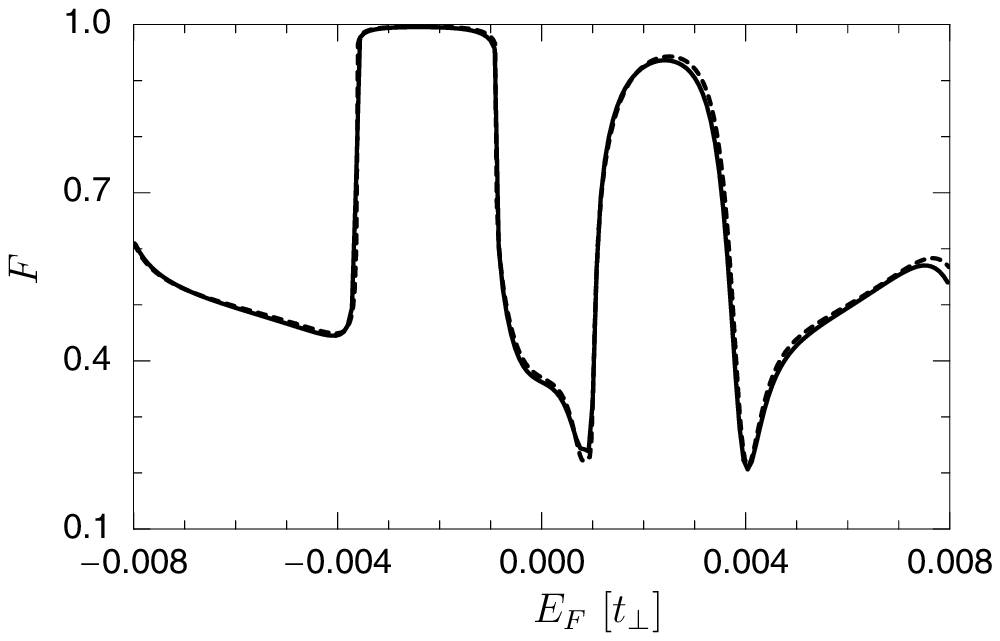}
    \caption{Conductivity (top) and Fano factor (bottom) around the Dirac point, for $L=100\, l_\perp$ and $U_\infty=0.2\,t_\perp$. (These parameter values correspond to the intersection point of our curves with the prediction of Ref.\ \onlinecite{Kat06b} in Fig.\ \ref{cfv}.) The solid lines were obtained using the four-band Hamiltonian (\ref{Hdef}), while the dashed lines were obtained from the two-band Hamiltonian (\ref{H2def}).\label{compare}}
  \end{center}
\end{figure}

We have repeated the calculation of conductance and Fano factor using both Hamiltonians (\ref{Hdef}) and (\ref{H2def}), for parameter values 
corresponding to the intersection point of Fig.\ \ref{cfv}, and find good agreement (see Fig.\ \ref{compare}).


\begin{thebibliography}{99}
\bibitem{Two06} J. Tworzyd{\l}o, B. Trauzettel, M. Titov, A. Rycerz, and C. W. J. Beenakker, Phys.\ Rev.\ Lett.\ {\bf 96}, 246802 (2006).
\bibitem{reviews} For a review and a tutorial on shot noise we refer, respectively, to:  Ya.\ M. Blanter and M. B\"{u}ttiker, Phys.\ Rep.\ {\bf 336}, 1 (2000); C. W. J. Beenakker and C. Sch\"{o}nenberger, Physics Today {\bf 56} (5), 37 (2003).
\bibitem{Wal47} P. R. Wallace, Phys.\ Rev.\ {\bf 71}, 622 (1947).
\bibitem{McCann06} E. McCann and V. I. Fal'ko, Phys.\ Rev.\ Lett.\ {\bf 96}, 086805 (2006).
\bibitem{Nil06b} J. Nilsson, A. H. Castro Neto, N. M. R. Peres, and F. Guinea, Phys.\ Rev.\ B {\bf 73}, 214418 (2006).
\bibitem{note1} This value of the interlayer coupling strength $\gamma_{1}$ refers to graphite; the value for a bilayer is not yet known.
\bibitem{Cse06} J. Cserti, cond-mat/0608219.
\bibitem{Kos06} M. Koshino and T. Ando, Phys.\ Rev.\ B {\bf 73}, 245403 (2006).
\bibitem{Nil06} J. Nilsson, A. H. Castro Neto, F. Guinea, and N. M. R. Peres, cond-mat/0604106.
\bibitem{Kat06} M. I. Katsnelson, Euro.\ Phys.\ J.\ B {\bf 51}, 157 (2006). 
\bibitem{Nov06} K. S. Novoselov, E. McCann, S. V. Morozov, V. I. Fal'ko, M. I. Katsnelson, U. Zeitler, D. Jiang, F. Schedin and A. K. Geim, Nature Physics {\bf 2}, 177 (2006).
\bibitem{Kat06b} M. I. Katsnelson, Euro.\ Phys.\ J.\ B {\bf 52}, 151 (2006). 
\bibitem{Nil06c} J. Nilsson, A. H. Castro Neto, F. Guinea, and N. M. R. Peres, cond-mat/0607343.
\bibitem{McC06} E. McCann, cond-mat/0608221.
\bibitem{Kat06c} M. I. Katsnelson, K. S. Novoselov, and A. K. Geim, Nature Physics {\bf 2}, 620 (2006).
\end{thebibliography}
\end{document}